\documentclass[cits]{PoS}

\usepackage{amsmath,amssymb}

\newcommand{\be}{\begin{equation}}
\newcommand{\ee}{\end{equation}}

\newcommand{\ket}[1]{|\,#1\,\rangle}

\newcommand{\ud}{\mathrm{d}}

\newcommand{\E}{\mathbf{E}}
\newcommand{\B}{\mathbf{B}}
\newcommand{\vv}{\mathbf{v}}
\newcommand{\pp}{\mathbf{p}}
\newcommand{\A}{\mathbf{A}}

\newcommand{\x}{\mathbf{x}}

\newcommand{\z}{\mathbf{z}}

\newcommand{\avec}{\mathbf{a}}

\title{Physical charges in QED and QCD}

\ShortTitle{Physical charges in QED and QCD}

\author{\speaker{Anton Ilderton}\thanks{Address from November 2009: Department of Physics, Ume\aa\ University, Ume\aa, Sweden.}
\\
        Trinity College, Dublin\\
        E-mail: \email{antoni@maths.tcd.ie}}

\author{Martin Lavelle\\
        School of Computing and Mathematics, University of Plymouth\\
        E-mail: \email{martin.lavelle@plymouth.ac.uk}}

\author{David McMullan\\
        School of Computing and Mathematics, University of Plymouth\\
        E-mail: \email{david.mcmullan@plymouth.ac.uk}}

\abstract{
We show that the `dressing' approach, which describes physical charges as gauge invariant composites of matter and clouds of gauge bosons, arises naturally in gauge theories. We give perturbative examples
of dressings for both asymptotic charges and for states in which the fields are
confined to a compact volume as is required, for example, by causality in pair
creation. In QCD, we use dressed states to demonstrate explicitly how Gribov copies obstruct the
non-perturbative construction of colour charges.
}

\FullConference{International Workshop on QCD Green's Functions, Confinement
and Phenomenology\\
                 September 7-11 , 2009\\
                 ECT Trento, Italy}

\begin{document}

\section{Introduction}

The infra-red problems which plague gauge theories can be traced back to the assumption that the coupling switches off asymptotically, so that the Lagrangian become free and can be identified with physical particles. While this may be an acceptable in Yukawa type theories, it clearly does not hold in QED or QCD -- the latter giving the most obvious counterexample, since there are no free quarks or gluons \cite{Kulish:1970ut, Horan:1999ba}.

The essential idea behind the dressing approach is that in order to describe physical charges we must consider not just the Lagrangian fermions fields $\psi$, but composite operators $h^{-1}[A]\psi$, where the `dressing' $h^{-1}[A]$ transforms as
\begin{equation*}
	h^{-1}[A] \to h^{-1}[A] U\qquad\text{when}\qquad \psi\to U^{-1}\psi\;.
\end{equation*}
This is of course a minimal requirement, such that the composite is gauge invariant. We will discuss explicit forms of the dressing below. The dressing approach has appeared in various guises in the literature, and has been used for a variety of purposes, see for example \cite{Cornwall:1979hz, Prokhorov:1991fs, Lunev:1994ty}. Here we will take a bottom-up approach, starting with a simple gauge theory which can be solved exactly, and showing that physical states are indeed dressed states. For example, we will construct dressed states which describe asymptotic, or ground state, charges. We will then construct states in which the electromagnetic fields are confined to a compact region. These are necessary for treating processes such as pair creation and hadronisation, where causality constrains the spatial extent of the fields.

Finally, we turn to non-abelian theories and discuss the nonperturbative impact of the Gribov ambiguity on the construction of coloured charges.

\section{Physical charges}
For concreteness and simplicity we focus in this section on the toy model of U(1) gauge fields with heavy matter, fixed to move at velocity $v^\mu$, i.e.
\be\label{act}
	\mathcal{L} = -\frac{1}{4}F_{\mu\nu}F^{\mu\nu} + i\,Q^\dagger v^\mu \big(\partial_\mu +i e A_\mu\big)Q\;.
\ee
This is just the leading order, in $m^{-1}$, `HQET' action for QED \cite{Georgi:1990um}. The use and renormalisation of our dressed charges in perturbation theory is discussed thoroughly in \cite{Bagan:1999jf, Bagan:1999jk}. Working in a Hamiltonian picture, we have two equations to solve in order to identify physical states, namely the Schr\"odinger equation and Gauss's law,
\be
	\hat H\ket\Psi = i\partial_t\ket\Psi\;,\qquad \nabla\!\cdot\!\E \ket\Psi = -eQ^\dagger Q\ket\Psi\;,
\ee
respectively. In this theory we can solve both, and so write down physical states, exactly. For example, the gauge invariant ground state wavefunctional of a single moving charge in the Schr\"odinger representation is, up to a phase which we do not need here \cite{Ilderton:2009jb},
\be\label{0state}
	\Psi[\A,Q^\dagger] = \exp\bigg[ ie\int\!\frac{\ud^3p}{(2\pi)^3}\ \mathrm{e}^{-i\pp\cdot\x_t}\bigg(\frac{i\pp}{|\pp|^2}-\frac{i\vv^T}{|\pp|^2-(\vv.\pp)^2}\bigg)\cdot \A(\pp)\bigg]Q^\dagger(\x_t)\Psi_0[\A^T]\;,
\ee
where $\Psi_0$ is the gauge invariant photon vacuum, $\x_t:=\x_0+\vv t$ is the path along which the heavy charge moves, and a superscript $T$ denotes projection onto the transverse part using the operator $T_{ij}=\delta_{ij}-p_ip_j/\pp^2$.

This is a `dressed' state: the matter field $Q^\dagger$ is surrounded by a cloud of `photons', and neither part is individually gauge invariant: only together do they give a gauge invariant  charge. The longitudinal term, in $\pp\cdot\A$, is required by Gauss's law and this is what ensures gauge invariance of the state\footnote{It is easily checked that the state (\ref{0state}) is invariant under $Q^\dagger\to \exp(ie\Lambda)Q^\dagger$ and $\A\to \A-\nabla\Lambda$.}. Solving the Schr\"odinger equation generates the gauge invariant terms in $\vv^T\cdot\A$ which ensure the correct physics. If we calculate the electromagnetic fields of this state, we find
\begin{eqnarray}
	\label{elec} \langle\E(\z,t)\rangle_\Psi &=& -\frac{e\gamma}{4\pi}\frac{\z-\x_t}{\big[|\z-\x_t|^2+\gamma^2(\vv.(\z-\x_t))^2\big]^{3/2}}\;, \\[5pt]
	\label{mag} \langle\B(\z,t) \rangle_\Psi &=& \vv\times \langle\E(\z,t)\rangle_\Psi\;,
\end{eqnarray}
where $\gamma\equiv\sqrt{1+\vv^2}$ is the usual Lorentz gamma factor. These are precisely the electromagnetic fields of a charged particle moving along the path $\x_t$ \cite{Jackson}.

Note that at $\vv=0$, the state (\ref{0state}) describes Dirac's static electron \cite{Dirac:1955uv}: the exponent of the wavefunctional reduces to
\be\label{stat0}
	ie\frac{\nabla\!\cdot\!\A(\x_0)}{\nabla^2} \equiv -\frac{ie}{4\pi} \int\limits\!\ud^3\x\ \nabla\bigg(\frac{1}{|\x-\x_0|}\bigg) \cdot \A(\x)\;,
\ee
where we have inverted the Fourier transform and, on the right, used the explicit form for $\nabla^{-2}$. The magnetic field (\ref{mag}) vanishes in the static limit and (\ref{elec}) reduces to the Coulombic electric field of a static charge.

These states are clearly nonlocal, which is an immediate consequence of Gauss's law. This nonlocality is inevitable in gauge theories; for example, recall that, in QED, the charge operator acting on a gauge invariant state takes the form
\be
	{\mathrm{Q}}\ket\Psi = \int\limits_{S_\infty}\!\ud\mathbf{s} \!\cdot\! \E\ket\Psi\;,
\ee
which is the radial electric field component integrated over the sphere at infinity. If this is to return a non--zero value, the charge must be nonlocal. Nevertheless, we saw in (\ref{elec}) and (\ref{mag}) that the state has good, local observables.  Our states were derived simply from imposing Gauss's law and solving the Schr\"odinger equation, so we see that dressed charges arise very naturally in gauge theory. The calculations presented here can of course be extended to QED proper, and QCD in perturbation theory \cite{Bagan:1999jf, Bagan:1999jk}.

Note that the fields (\ref{elec}) and (\ref{mag}) are spread through the whole of space. This is a direct consequence of Gauss's law. The same is true for the ground states of multiple charges described by adding more matter fields and identical dressing factors to the state (\ref{0state}). Such ground states therefore correspond to collections of {\it asymptotic} particles, the fields of which have had time to spread through all of space. In QED, it is indeed these dressed charges which become free at asymptotic times, whereas the `undressed' Lagrangian matter fields do not \cite{Kulish:1970ut, Horan:1999ba, Bagan:2000mk}.

Now, consider an $e^+$ $e^-$ pair created in some collision. At a finite time after creation, their electromagnetic fields must be confined to a finite volume of space, by causality. As such, Gauss's law must allow a description of charge neutral states in which the fields are constrained to a finite volume. We turn to such states below.

\section{Compact charges}
\begin{figure}[t!]
	\includegraphics[width=0.4\textwidth]{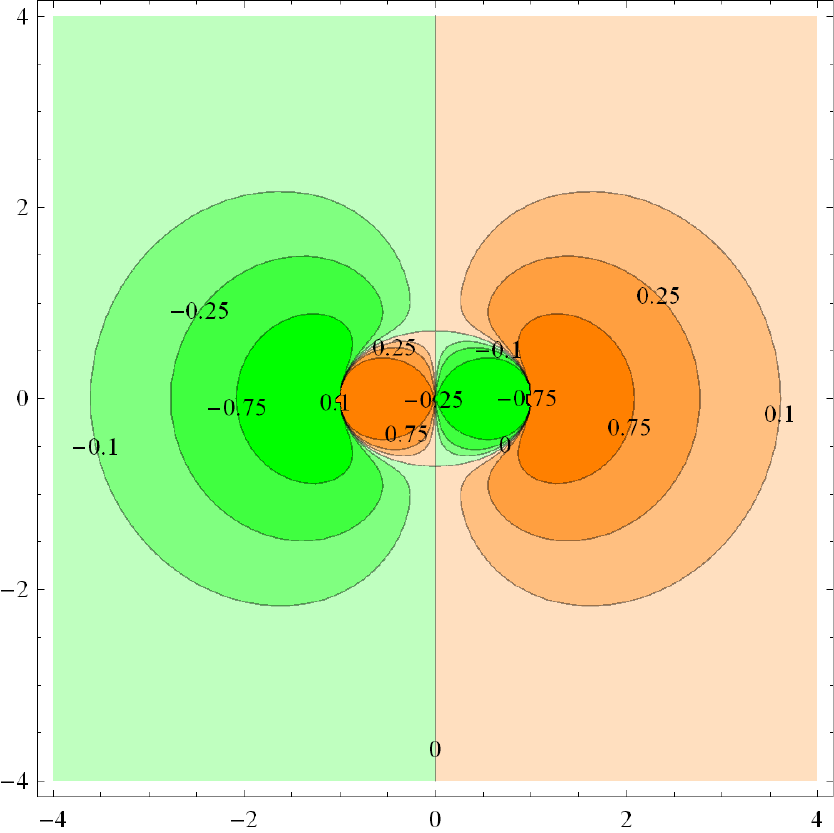}\hspace{20pt}\centering\includegraphics[width=0.4\textwidth]{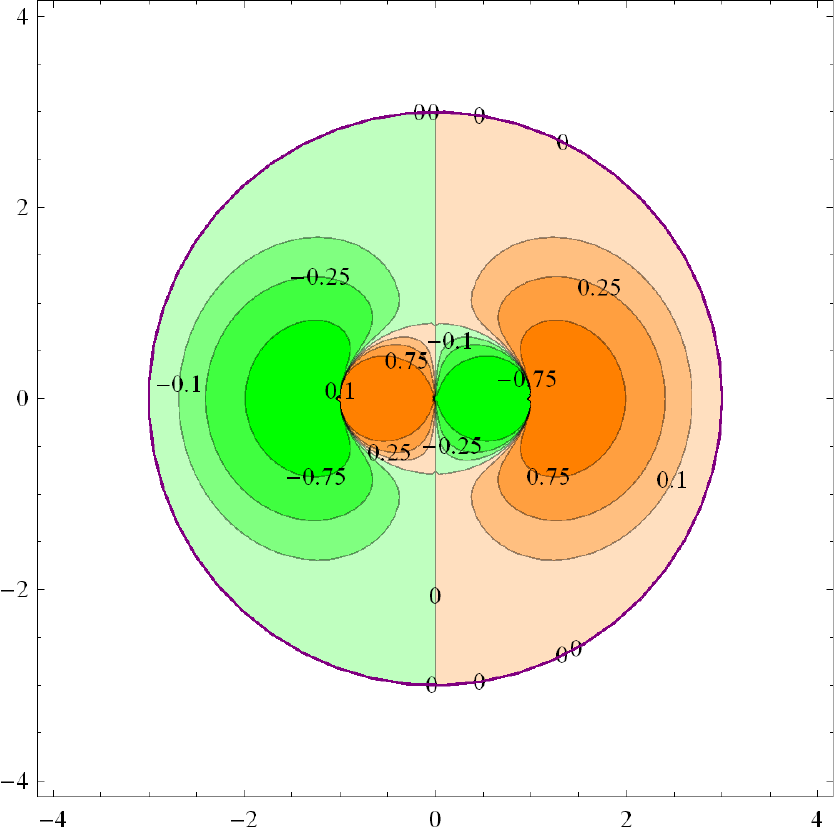}
	\caption{{\it Left:} the radial electric field of two asymptotic, static charges. {\it Right:} the radial electric field of two compact charges. Note the deformation of the field due to its restriction to finite volume, and the vanishing field strength on the bounding sphere.}
	\label{elecpic}
\end{figure}

For simplicity, consider two static charges at positions $\pm\mathbf{a}$; we will return to moving charges shortly. From (\ref{stat0}), the exponent of the dressing for two asymptotic (ground state) charges of opposite sign and positions $\pm\avec$ is
\be\label{one}
	\frac{1}{4\pi}\int\limits\!\ud^3\x\ \nabla\bigg(\frac{1}{|\x-\avec|}-\frac{1}{|\x+\avec|}\bigg)\cdot \A(\x)\;,
\ee
which describes two of Dirac's static electrons. Choosing a volume $V$ in which the electromagnetic fields are to be confined, the appropriate dressing is \cite{Ilderton:2009jb},
\be\label{two}
	\frac{1}{4\pi}\int\limits_{ V}\!\ud^3\x\ \nabla\bigg(\frac{1}{|\x-\avec|}-\frac{1}{|\x+\avec|} + {  f_{V} }(\x)\bigg)\cdot \A(\x)\;,
\ee
where we have restricted the spatial integration in (\ref{one}) to $V$, and introduced a function $f_V$ under the derivative. This function is determined, once $V$ is chosen, by demanding gauge invariance of the state. For example, in the case that $V$ is a sphere of radius $R$ centred on the origin, with $R>a\equiv|\avec|$, the function $f_V$ can be shown to be \cite{Ilderton:2009jb}
\begin{equation}\label{fv}
    f_V(r,\theta)=\frac4R \sum_{\ell=0}^\infty \left(\frac{\ell+1}{2\ell+1}\right) \left(\frac{a}{R}\right)^{2\ell+1} \left(\frac{r}{R}\right)^{2\ell+1} P_{2\ell+1}(\cos\theta)\;.
\end{equation}
This $f_V$ is independent of the azimuthal angle around the axis on which the two charges sit, and the $P_\ell$ are Legendre Polynomials. The resulting radial electric field component of the state is plotted in the right hand panel of Fig.~\ref{elecpic}. Comparing with the field of two asymptotic static charges, plotted in the left hand panel, the distortion of the compact field close to the boundary can be clearly seen. In particular, the radial field strength vanishes on the bounding sphere, as the total charge within the ball is zero.

We may also calculate the potential between the two charges, which is the analogue of the inter--quark potential. Before doing so, recall that we can make a gauge invariant charge--anticharge state by linking the matter fields with a Wilson line.  In this case, the electric field is also restricted to a compact region; it is only non--zero on the path connecting the charges. The potential between the charges in this state is linear rising, but with a divergent coefficient going like $\delta^2(0)$, due to the infinitesimal extent of the string in its transverse directions. So although this state is gauge invariant, it is infinitely excited and therefore completely unphysical. With this in mind, the potential $V_{cc}$ between our two compact charges can be found to be
\be\label{vcc}
	V_{cc} = -\frac{e^2}{4\pi}\frac{1}{2a}+\frac{e^2}{2\pi}\bigg(\frac{a^2R}{R^4-a^4}+\frac{1}{R}\tanh^{-1}\frac{a^2}{R^2}\bigg)\;,
\ee
\begin{figure}[!ht]
\centering\includegraphics[width=0.45\textwidth]{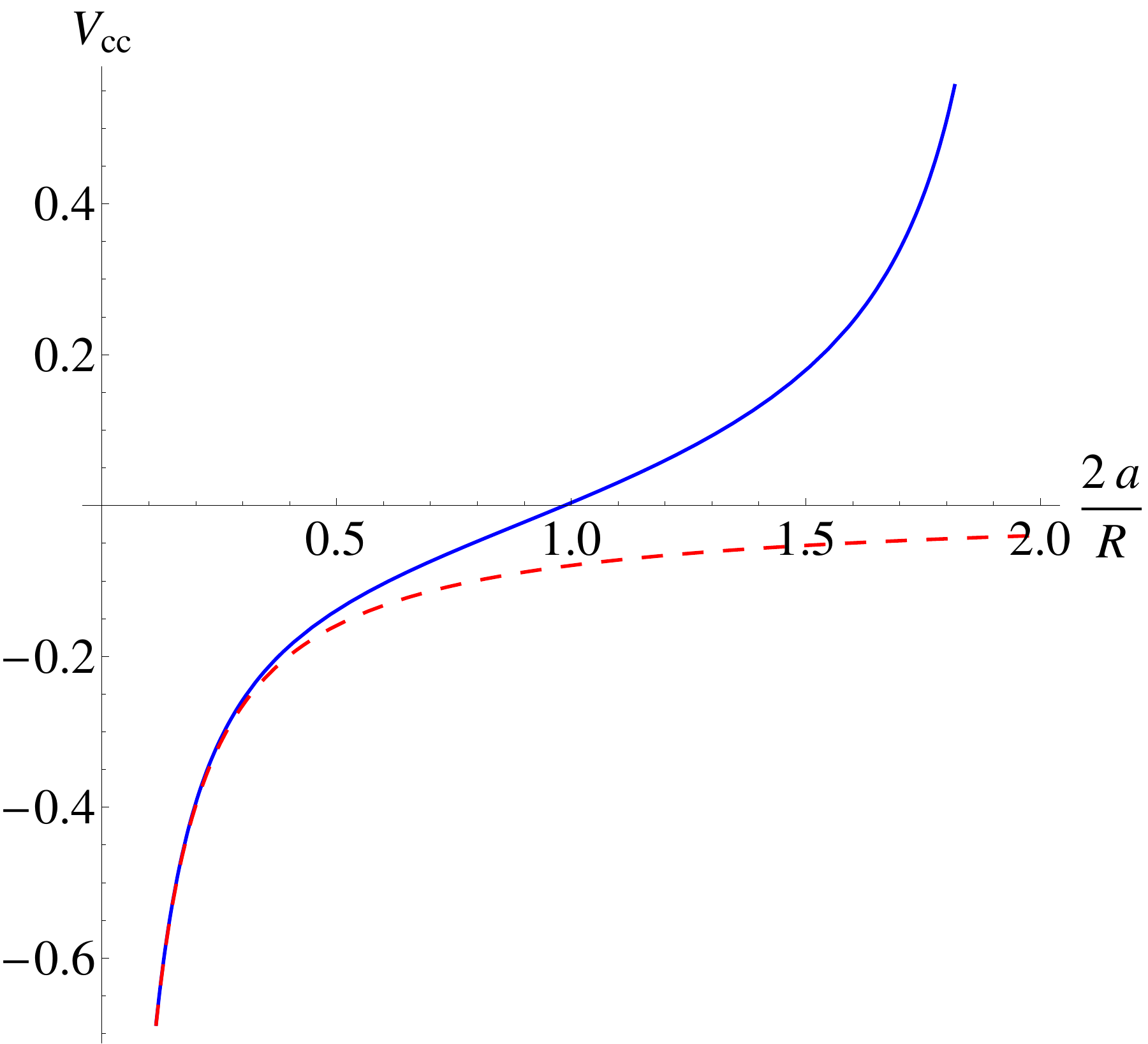}
\caption{{\it Blue (solid) line}: the potential $V_{cc}$ (in units of $1/R$) between two compact charges, plotted against their separation, relative to the radius of the ball, $2a/R$. {\it Dashed (red) line}: the Coulombic potential between two asymptotic charges, for comparison.}
\label{potplot}
\end{figure}
which is plotted in Fig.~{\ref{potplot}}. The first term of (\ref{vcc}) is just the Coulombic potential between the two charges at $\pm\avec$, and this is also plotted in Fig.~{\ref{potplot}} for comparison. Our inter--charge potential, unlike that in the Wilson line state, is finite for all separations of the charges $a<R$. (As it is clearly unphysical to pull the charges past the boundary of the field envelope, the potential diverges at $a=R$.) When the charges sit deep inside the ball, $a\ll R$, the potential is approximately Coulombic. This is just the statement that the charges want to be `fat'; i.e. their fields naturally expand out into the whole of space. This is what happens when time evolves; the charges throw off energy, and $V$ expands as the charges relax to their Coulombic ground state.

In this static charge system, the choice of $V$ is arbitrary, when considered as an initial condition for states in a fixed time slice. More generally, though, we would expect $V$ to be determined dynamically. For an example of this, we return to the case $\vv\not=0$. By considering states in which the charges occupy the same spatial position at some time, we have built models of pair creation and annihilation \cite{Ilderton:2009jb}. In the former, the electromagnetic fields propagate out from the point of creation  and are constrained to a volume which we have shown to be a ball at leading order in $t$, in agreement with causal expectations. In pair annihilation, the electromagnetic fields which remain after the matter fields have vanished continue to evolve under the free photon Hamiltonian in such a way that a cavity forms around the annihilation point. This cavity eats up the fields as it expands, until eventually all observers agree that the charges have vanished, and the state decays to the empty photon vacuum.

\section{Coloured charges}
We now turn to the non-abelian theory, and consider the dressings which describe gauge invariant colour charges, i.e. quarks. Recall that in order for the composite object $q:=h^{-1}[A]\psi$ to be gauge invariant if we transform $\psi\to U^{-1}\psi$, the dressing must obey
\be
	h^{-1}[A^U] = h^{-1}[A]U\;,\qquad A^U_\mu \equiv U^{-1}A_\mu U +\frac{1}{g}U^{-1}\partial_\mu U\;,
\ee
To proceed, we need to understand where objects like $h^{-1}$ come from. To pursue this, consider some gauge fixing condition, call it $\chi(A)=0$. Now, given any gauge field $A$, we find the field--dependent transformation $h[A]$ such that the field is taken into the chosen gauge, i.e. $\chi(A^{h[A]})=0$. We then do the same thing starting with the field at any other point on its gauge orbit $A^U$, i.e. we find $h[A^U]$ such that $\chi(A^{Uh[A^U]})=0$. Provided we have a good gauge fixing condition, the two fields in the gauge slice must be equal. From this we read off the transformation property of $h[A]$,
\begin{equation}
	A^{h[A]} = A^{U h[A^U]} \implies h^{-1}[A^U] = h^{-1}[A]U\;.
\end{equation}
In other words, dressings for single charges are the, inverses of, field--dependent gauge transformations which take us into particular gauges. Note that if we reverse these arguments, then, starting from a given dressing, we can construct from it a gauge fixing condition. To illustrate these ideas, we can generate the `Coulomb dressing' by solving the condition
\be
	\nabla\cdot\big(h^{-1}\A h + \frac{1}{g}h^{-1}\nabla U\big)=0\;,
\ee
for $h^{-1}$ as a function of $\A$. This can be performed in perturbation theory, and generalises Dirac's static electron to the non-abelian theory \cite{Lavelle:1995ty}.

Perturbatively, therefore, we are able to write down gauge invariant quarks, but we know there should be a nonperturbative obstruction to doing so, since no free quarks are observed in nature. Note that the construction above relied on having a good gauge fixing condition. Nonperturbatively, however, there is no perfect gauge fixing. All admissible conditions $\chi$, in the sense that they can be imposed on the space of gauge fields which fall appropriately quickly at infinity \cite{Singer:1978dk}, have Gribov copies \cite{Gribov:1977wm, vanBaal:1991zw, Ilderton:2007qy}. The impact of the Gribov ambiguity on the construction of a coloured charge is seen by returning to $h[A]$. Due to the copies, the field--dependent gauge transformation $h[A]$ is multiply defined, since there are multiple configurations obeying $\chi(A)=0$ along a particular orbit, and therefore $h^{-1}[A]$ does not exist. Gribov copies, therefore, prevent us from constructing a colour charged state nonperturbatively.

Now, if one can excise Gribov copies from the theory, then a gauge invariant colour charge can be defined (and from it, a perfect gauge fixing condition). For example, time evolution is discontinuous due to the presence of copies, at least in Coulomb gauge \cite{Jackiw:1977ng}, and so one could consider restricting the space of gauge fields to the fundamental modular region. Even then, however, Gribov copies remain on the boundary. In this case one has to identify state wavefunctionals in order to have a well--defined time evolution. If the copies can be removed in this way, or dealt with using some other approach, see for example \cite{Heinzl:2007cp, vonSmekal:2007ns, vonSmekal:2008es, Mehta:2009zv}, then we must look elsewhere for the confinement mechanism. If, however, one {\it cannot} remove the copies entirely, then physical states must be white, i.e. colour must be confined.

\section{Conclusions}

We have seen that the `dressed' description of charges as composites of matter fields and boson clouds arises naturally in gauge theories. We have given the ground state (asymptotic) dressing for a moving charge, and using these states to describe asymptotic particles, rather than assuming that the matter fields become free, is known to vastly improve the infra--red behaviour of scattering amplitudes \cite{MLTalk}. We have also given, for the first time, a description of a pair of compactly dressed charges, in which the fields are constrained to a finite, but non-singular, volume. Unlike the gauge invariant, but unphysical, state described by linking two matter fields with a Wilson line, the potential between the compact charges is finite.

In future work we will extend this construction to non-abelian theories, which will allow us to model mesonic states. It would also be extremely interesting to relate the inherent nonlocality of our charges to the nonlocality of the horizon function appearing in the Gribov--Zwanziger Lagrangian \cite{GraceyTalk} and related nonlocal operators \cite{Gracey:2007ki}.

\acknowledgments
A.~I. kindly thanks the organisers of QCD-TNT for the opportunity to speak, and R.~ Ferrari, J.~A.~Gracey, V.~P.~Nair and A.~Szczepaniak for useful discussions. A.~I. is supported by IRCSET.

\end{document}